\documentclass{iaus}
\usepackage{graphicx}

\title[A LAMOST Spectroscopic Survey of the Galactic Anti-center] 
{LSS-GAC -- A LAMOST Spectroscopic Survey of the Galactic Anti-center}

\author[X.-W. Liu et al.]   
{X.-W. Liu$^{1,2}$, 
H.-B. Yuan$^2$, Z.-Y. Huo$^3$, 
L.-C. Deng$^3$, J.-L. Hou$^4$, Y.-H. Zhao$^3$,
G. Zhao$^3$, J.-R. Shi$^3$ \and A.-L. Luo$^3$, 
M.-S. Xiang$^1$, 
H.-H. Zhang$^1$, Y. Huang$^1$, 
H.-W. Zhang$^1$
 \thanks{On behalf of the LAMOST Science Committee,
  LAMOST Scientific Performance 
  Working Group
  and the LAMOST Galactic Surveys Working Group.
}}

\affiliation{$^1$ Department of Astronomy, Peking University, 
Beijing, P.\,R.\,China\\ email: {\tt x.liu@pku.edu.cn} \\[\affilskip]
$^2$ Kavli Institute for Astronomy and Astrophysics, Peking University,
Beijing, P.\,R.\,China  \\[\affilskip]
$^3$ National Astronomical Observatories, Chinese Academy of Sciences,
Beijing, P.\,R.\,China \\[\affilskip]
$^4$ Shanghai Astronomical Observatory, Chinese Academy of Sciences,
Shanghai, P.\,R.\,China 
}

\pubyear{2014}
\volume{298}  
\pagerange{XX -- YY}
\jname{IAUS\,298 Setting the scene for Gaia and LAMOST}
\editors{S. Feltzing, G. Zhao, N.\,A. Walton \& P.\,A. Whitelock, eds.}
\begin{document}

\maketitle

\begin{abstract}

As a major component of the LAMOST Galactic surveys, the LAMOST Spectroscopic
Survey of the Galactic Anti-center (LSS-GAC) will survey a significant volume
of the Galactic thin/thick disks and halo in a contiguous sky area of
$\sim3,400$\,sq.deg., centered on the Galactic anti-center ($|b| \leq
30^{\circ}$, $150 \leq l \leq 210^{\circ}$), and obtain
$\lambda\lambda$3800--9000 low resolution ($R \sim 1,800$) spectra for a
statistically complete sample of $\gtrsim 3$\,M stars of all colors, uniformly
and randomly selected from ($r$, $g - r$) and ($r$, $r - i$) Hess diagrams
obtained from a CCD imaging photometric survey of $\sim 5,400$\,sq.deg. with
the Xuyi 1.04/1.20\,m Schmidt Telescope, ranging from $r = 14.0$ to a limiting
magnitude of $r = 17.8$ (18.5 for limited fields). The survey will deliver
spectral classification, radial velocity $V_{\rm r}$ and stellar parameters
(effective temperature $T_{\rm eff}$, surface gravity $\log\,g$ and metallicity
[Fe/H]) for millions of Galactic stars. Together with Gaia which will provide
accurate distances and tangential velocities for a billion stars, the LSS-GAC
will yield a unique dataset to study the stellar populations, chemical
composition, kinematics and structure of the disks and their interface with the
halo, identify streams of debris of tidally disrupted dwarf galaxies and
clusters, probe the gravitational potential and dark matter distribution, map
the 3D distribution of interstellar dust extinction, search for rare objects
(e.g.  extremely metal-poor or hyper-velocity stars), and ultimately advance
our understanding of the assemblage of the Milky Way and other galaxies and the
origin of regularity and diversity of their properties.

The survey was initiated in the fall of 2012 and expected to complete in the
spring of 2017.  Hitherto, about 0.4\,M spectra of ${\rm S/N}(\lambda7450) \ge
10$ per pixel have been accumulated. In addition, bright nights have been used
to target stars brighter than 14 mag.\ and have so far generated over 0.4\,M
spectra, yielding an excellent sample of local stars to probe the solar
neighborhood. LSP3, a set of pipelines tailored to the need of LSS-GAC, for
spectral flux-calibration, and radial velocity and stellar parameter
determinations, have been developed at Peking University, based on packages
developed for the SDSS and those at the National Astronomical Observatories of
Chinese Academy of Sciences. Comparisons of multi-epoch observations, with the
SDSS results, as well as applying the pipelines to open and globular clusters
show that LSP3 has achieved a precision of 5\,km\,s$^{-1}$, 110\,K, 0.15\,dex
and 0.15\,dex for $V_{\rm r}$, $T_{\rm eff}$, $\log\,g$ and [Fe/H],
respectively.  The data are publicly available, subject to regulations of the
LAMOST data policy, and begin to yield scientific results. The potential of
LSS-GAC and power of existing data are illustrated with examples of selected
problems.

\end{abstract}

\firstsection 
\section{Scientific motivations}

Galaxies are building blocks of the universe. Understanding how galaxies
assemble and acquire their characteristic structure and properties (the ``grand
designs'') is one of the most challenging problems of astrophysics. In the
current $\Lambda$CDM paradigm, all cosmic structures including galaxies build
up hierarchically by merging and accreting smaller ones. Although supported by
extensive observations and large scale numerical simulations, the scenario
involves many currently poorly understood and often highly non-linear baryonic
physics (e.g. star formation and feedbacks) that it is compulsory to find out
empirically the assemblage history of galaxies via detailed study of large
samples of member stars of nearby, well-resolved galaxies, and inevitably less
sophisticated statistical analysis of large samples of more distant galaxies
(e.g. Mayer, Governato \& Kaufmann 2008). Both approaches depend on the
acquisition of large survey samples.

The Milky Way is an archetypical disk galaxy and the only grand-design (barred)
spiral that individual constituent stars can be resolved and studied
multi-dimensionally (in 3D space position and velocity, and in chemical
composition). On the other hand, owing to our own position inside the system,
the roughly 200 billion Galactic stars are distributed over the whole sky of
$4\pi$ steradian, and their views are restricted by the ubiquitous intervening
interstellar dust grains, in particular in the disk and towards the Galactic
center. The study is further hampered by the difficulty of obtaining accurate
distances to individual stars. Owing to the horrendous technical challenges of
a complete spectroscopic census of the Galaxy, far beyond the capabilities of
all existing facilities, hitherto progress of Galactic spectroscopic surveys
lags far behind that of extragalactic, and as a consequence, our knowledge of
our own host galaxy (contents of baryonic and dark matter, kinematics and
chemical composition) is, ironically, not as complete and precise as that of
the universe as a whole.

Modern large scale surveys, represented by the extremely successful SDSS (York
et al. 2000), have revolutionized our understanding of galaxy formation and
evolution, including the still on-going assemblage process of the Milky Way. On
the other hand, being primarily an extragalactic program, the volume of the
Galactic disk, in particular the thin disk, sampled by the SDSS, including its
Galactic extensions SEGUE (Yanny et al. 2009) and APOGEE (Eisenstein et al.
2011; M. Schultheis, this volume), is very limited. Of the 0.7\,M stellar
spectra released in the SDSS DR9 (Ahn et al. 2012), a significant number were
secured as the auxiliary data (e.g. as flux or radial velocity standard stars),
with the remaining distributed in hundreds of disconnected fields, with targets
selected from a host of algorithms. It is thus difficult to carry meaningful
statistical analyses in terms of the underlying stellar populations.  The
operation of the Chinese LAMOST (Cui et al. 2012; Deng et al., this volume)
opens up the possibility of carrying out a systematic spectroscopic survey that
samples a significant volume of the Milky Way, in particular the disk, the
defining structure that contains more than 90\% of the baryonic matter and
essentially all the angular momentum of the Galaxy.  Compared to the facility
of SDSS, the biggest strength of LAMOST is its sheer number of fibers -- 4,000,
which is respectively a factor of 6.25 and 4 larger than that employed in the
SDSS and SDSS-II, and in the SDSS-III. Most clear nights of the Xinglong
Station where the LAMOST is located are distributed from September to March of
the coming year, peaking in December when the Galactic anti-center (${\rm RA} =
5^{\rm h}46^{\rm m}$, ${\rm Dec} = +28^{\circ}56^{\prime}$) culminates around
midnight. The outer disk also suffers less severe dust extinction. Most
important of all, the outer parts of the Galactic disk have already revealed
complex structure that is poorly understood. LSS-GAC, a LAMOST Spectroscopic
Survey of the Galactic Anti-center and a major component of the LAMOST
Experiment for Galactic Understanding and Exploration (LEGUE) proposed by the
LAMOST Galactic Surveys Working Group, thus offers an unique opportunity to
address questions key to revealing the true multi-dimensional structure and the
formation and evolution history of the Galactic disk, and of the Galaxy as a
whole. Examples of selected ``puzzles'' that can be tackled with the LSS-GAC
include disk formation of the Milky Way and other late type spirals, secular
evolution and (in)stability to gravitational perturbations of the disk, origin
of the thick disk, structures and substructures of the outer disk [truncation,
warps and flares, the Monoceros Ring and other anti-center stellar
(sub)structures].  

The LSS-GAC will form excellent synergy with Gaia, the forthcoming next
generation space all-sky astrometric survey (Perryman et al. 2001; A.
Vallenari, this volume). With an unprecedented astrometric accuracy of 10
micro-arcsec, Gaia will provide accurate parallaxes and proper motions for a
billion stars. Although the Gaia/RVS spectrometer will also provide radial
velocities and elemental abundances for millions of stars, they are limited to
stars brighter than 15 -- 16 mag., about 2 -- 3 mag.\, shallower than
achievable with the LAMOST. Radial velocities and basic stellar parameters
(effective temperatures, surface gravities and metallicities) of millions of
Galactic stars derivable from LAMOST spectra, together with accurate distances
and tangential velocities yielded by Gaia will allow one to study the stellar
populations, kinematics and the star formation and chemical enrichment history
of a significant volume of the Milky Way in unprecedented detail. 

\section{Survey design}

The LSS-GAC will survey a contiguous sky area of $\sim 3,400$\,sq.deg.,
covering Galactic longitudes $150 \leq l \leq 210^{\circ}$ and latitudes $|b|
\leq 30^{\circ}$ (Fig.\,1), and obtain $\lambda\lambda$3800--9000 low
resolution ($R \sim 1,800$) spectra for a statistically complete sample of
$\gtrsim 3$\,M stars of all colors, uniformly and randomly selected from ($r$,
$g - r$) and ($r$, $r - i$) Hess diagrams and in (RA, Dec) spatial
distribution, obtained from the XSTPS-GAC, a CCD imaging photometric survey of
$\sim 5,400$\,sq.deg. using the Xuyi 1.04/1.20\,m Schmidt Telescope. In order
to make efficient use of observing time of different qualities and avoid fiber
crosstalk, three categories of spectroscopic plates are designated, {\bf
B}right, {\bf M}edium-bright and {\bf F}aint, targeting sources of brightness
$14.0 < r \lesssim m_1$, $m_1 \lesssim r \lesssim m_2$ and $m_2 \lesssim r \leq
18.5$, respectively.  The border magnitude separating B and M plates, $m_1$,
and that separating M and F plates, $m_2$, have values about 16.3 and 17.8,
respectively, but the exact numbers vary slightly from field to field, owing to
the varying source density, stellar population and interstellar extinction. In
general, $m_1$ and $m_2$ decrease towards the Galactic plane. Variable $m_1$
and $m_2$ are necessary in order to avoid discontinuities in the magnitude
distributions of stars targeted by B, M and F plates, c.f. {\S{3} for detail.
Also note that during the early stage of the pilot survey, October 2011 --
November 2011, the faint limit of F plates was set at 19.0 instead of 18.5. 

\begin{figure}[t]
\begin{center}
 \includegraphics[width=4in]{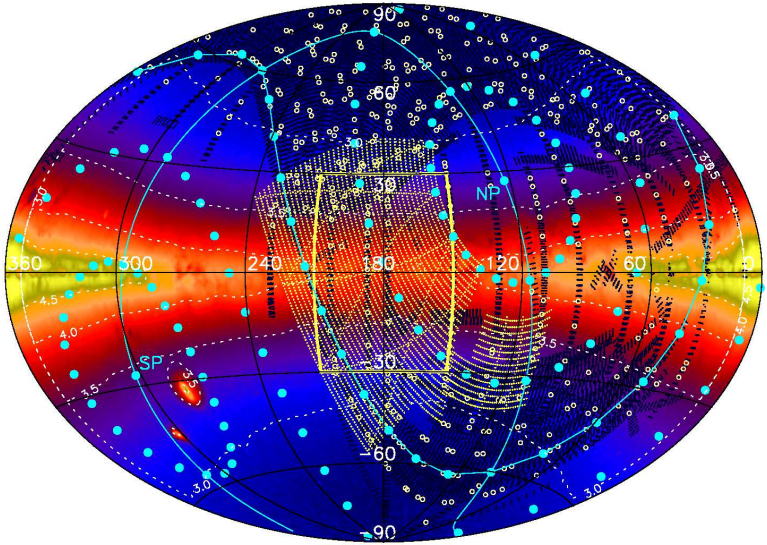} 

 \caption{Footprints of the LSS-GAC (the central yellow bucket box) and
XSTPS-GAC (the central yellow shaded area) in Galactic coordinates centered on
the Galactic anti-center ($l = 180^{\circ}$, $b = 0^{\circ}$). The background
pseudo color image shows the stellar number densities per squared degree from
the 2MASS survey (Skrutskie et al. 2006), overlaid with contours of constant
logarithmic number densities (the white dotted lines). The black shaded areas
delineate footprints of the SDSS and SDSS-II imaging surveys. White open
circles denotes SDSS/SEGUE spectroscopic plates. The cyan lines and dots
delineate the equatorial coordinate system, with the north and south celestial
poles marked.}

   \label{NNN:fig1} 
\end{center}
\end{figure}

The 4,000 fibers of LAMOST are distributed in a FoV of 20\,sq.deg., yielding a
fiber density of 200 per sq.deg. On average, $\sim 1,000$ stars will be
surveyed per sq.deg. outside the thin disk ($|b| > 3.5^{\circ}$), implying 5
plates (allocated as 2 B, 2 M and 1 F) per pointing position.  Within $|b| \leq
3.5^{\circ}$, the sampling is doubled. The LAMOST FoV however cannot be
arbitrarily arranged, since for the active optics to work in order to bring
individual segments of the primary and corrector mirrors into focus, there has
to be a star brighter than $\sim 8$\,mag. at the field center. Given this
constraint and considering that the LAMOST has a circular FoV, field centers
are carefully chosen, yielding overlapping FoV's, such that the uniformity of
spatial sampling over the whole survey area is maximized. The optimization is
carried out for B, M and F plates separately. In total, 1,250 pointings are
planned, including 500 B, 500 M and 250 F.
 
In the overlapping areas of adjacent FoV's, fibers are assigned to stars
regardless whether they have been assigned a fiber or not in the adjacent
plates. As a result, $\sim 23$\% of all selected targets will actually be
targeted twice, opening up the possibility of time-domain spectroscopy for,
e.g.  identifications of spectroscopic binaries. 

Finally, observing time under bright lunar conditions are used to target very
bright stars of magnitudes between $\sim 9$ and 14. They form an excellent
sample of local stars to probe the solar neighborhood. The VB plates are
implemented and scheduled separately. 

\section{XSTPS-GAC and target selection}

To provide input catalogs for the LSS-GAC, XSTPS-GAC -- the Xuyi Schmidt
Telescope Photometric Survey of the Galactic Anti-center was initiated in the
fall of 2009 and completed in the spring of 2011. The survey was carried out in
SDSS $g$-, $r$- and $i$-bands with the Xuyi 1.04/1.20\,m Schmidt Telescope
equipped with a 4k$\times$4k CCD camera, operated by the Near Earth Objects
Research Group of the Purple Mountain Observatory. The CCD offers a FoV of
$1.94^{\circ}\times 1.94^{\circ}$, with a pixel scale of 1.705\,arcsec. The
survey covers an area of $\sim 5,400$\,sq.deg., from ${\rm RA} \sim 3$ to
$9^{\rm h}$ and ${\rm Dec} \sim -10$ to $+60^{\circ}$, plus an extension of
$\sim 900$\,sq.deg. to the M\,31/M\,33 area. Including the bridging fields
connecting the two areas, the total survey area is close to 7,000\,sq.deg
(Fig.\,1).

The survey was carried out under good observing conditions (but not necessarily
photometric) in grey and dark nights. In a given night, fields of two adjacent
stripes of constant declinations were observed in turn, stepping the field
center in RA by half the FoV ($0.95^\circ$, i.e. $\sim 50$\% overlap). The
stepping in Dec between the stripes was set at 1.9$^{\circ}$, yielding an
overlap of $\sim 2$\,arcmin between two adjacent stripes. With an exposure time
of 90\,sec and a dual-channel readout time of 43.2\,sec, the moving of
telescope pointing direction as the observation proceeded was minimized,
creating maximum uniformity amongst the fields. To facilitate tying all the
fields to a common flux level (``ubercal''), ``Z-stripe'' fields that straddled
the borders of adjacent ``normal'' stripes were added. The images were
flat-fielded using ``super-sky flat fields'' generated from the target images
themselves, after clipping all visible stars. Astrometric calibration was
carried out using the PPMXL catalog (Roeser et al. 2010).  Aperture and PSF
photometric measurements were made using a modified DAOPHOT-based pipeline
developed by the BATC group of National Astronomical Observatories of Chinese
Academy of Sciences (NAOC), and then globally calibrated against the SDSS DR8
using overlapping sky areas. In total, the XSTPS-GAC archives approximately
100\,M stars down to a limiting magnitude of $\sim 19$ (10$\sigma$). The
astrometric accuracy is about 0.1\,arcsec, and a global photometric accuracy of
$\sim 2$\% has been achieved.

The LSS-GAC targets are selected from the XSTPS-GAC photometric catalogs. The
philosophy is to adopt a simple yet non-trivial algorithm, as uniform as
possible over the whole survey area, such that whatever objects (e.g.
white-dwarf-main-sequence binaries, extremely metal-poor or hyper-velocity
stars) are revealed by the spectroscopic observations, they can be studied in a
statistically meaningful way in terms of the underlying stellar populations (at
least for the survey volume), after various selection effects introduced by the
input catalog, those imposed by the constraints of instrument and observation
(e.g. the positioning of field center, the holes in focal plane produced by the
guiding CCDs and the central Shack Hartmann Sensor, allocations of fibers,
spectral S/N's, etc) have been properly taken into account.  Meanwhile in order
to increase the discovery space, rare objects of extreme colors (e.g.  white
dwarfs, red giants) should be preferentially targeted (Fig.\,2).

\begin{figure}[t]
\begin{center}
 \includegraphics[width=4in]{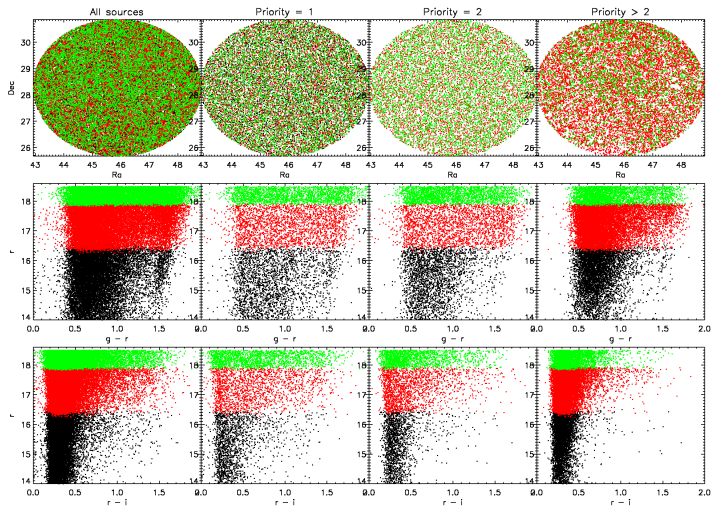} 

 \caption{From top to bottom, (RA, Dec) spatial distribution and ($r$, $g-r$)
and ($r$, $r-i$) Hess diagrams for all stars (first column), stars of first
(second column), second (third column) and lower (fourth column) priorities in
the clean sample, for a field around ${\rm RA} = 3^{\rm h}$ and ${\rm Dec} =
28^{\circ}$.}

   \label{NNN:fig2} 
\end{center}
\end{figure}

The XSTPS-GAC photometric catalogs are first used to generate a clean sample,
with the requirements that: 1) The stars are detected in at least two bands,
including $r$-band, and have $14.0 < r \leq 18.5$; 2) The star positions
measured in different bands agree within 0.5\,arcsec; 3) They are not flagged
as galaxies or star-pairs in either $r$- or $i$-band; 4) They have no neighbors
within 5\,arcsec that are brighter than $m + 1$\,mag., where $m$ is the
magnitude of star concerned, and are not neighbors of extremely bright stars.
 
To select targets, we first need to estimate values of $m_1$ and $m_2$, the
border magnitudes between B, M and F plates (c.f. \S{2} for detail).  For each
survey area in a (RA, Dec) box of one degree wide, stars of extremely blue
colors, $g - r$ or $r - i \leq -0.5$, and of extremely red colors, $g - r$ or
$r - i > 2.5$, are first selected.  Stars in the remaining color space are then
selected using a Monte Carlo method. First two random numbers, $r$ in the range
(14, 18.5] and $c$ in ($-1$, 5], are generated. If $-1 < c \leq 2$, then the
($r$, $g - r$) Hess diagram is used, assuming $g - r = c + 0.5$, otherwise the
($r$, $r - i$) Hess diagram is used instead, assuming $r - i = c
- 2.5$.  For each ($r$, $g - r$) or ($r$, $r - i$) set, if there are stars on
  the Hess diagram in a box centered on the simulated set of color-magnitude
and of length 0.2 in magnitude and 0.3 in color, then the star of
color-magnitude values closest to the simulated set is chosen and removed from
the pool. If not, the process is repeated until a total of 1,000 stars,
including those of extreme colors, per sq.deg. are selected. The stars are then
sorted in magnitude from bright to faint. Then $m_1$, the border magnitude
separating B and M plates is set to the faint end magnitude of the first 40\%
sources, and $m_2$ separating M and F plates is set to the faint end magnitude
of the first 80\% sources. 

The above procedure applies to the survey area of $|b| > 3.5^{\circ}$, where
the LSS-GAC plans to sample 1,000 stars per sq.deg. For $|b| \leq 3.5^{\circ}$,
the target density is doubled and 2,000 stars per sq.deg.  are selected. In
addition, in this case the selections are carried out in ($l$, $b$) instead of
(RA, Dec).

\begin{figure}[t]
\begin{center}
 \includegraphics[width=4in]{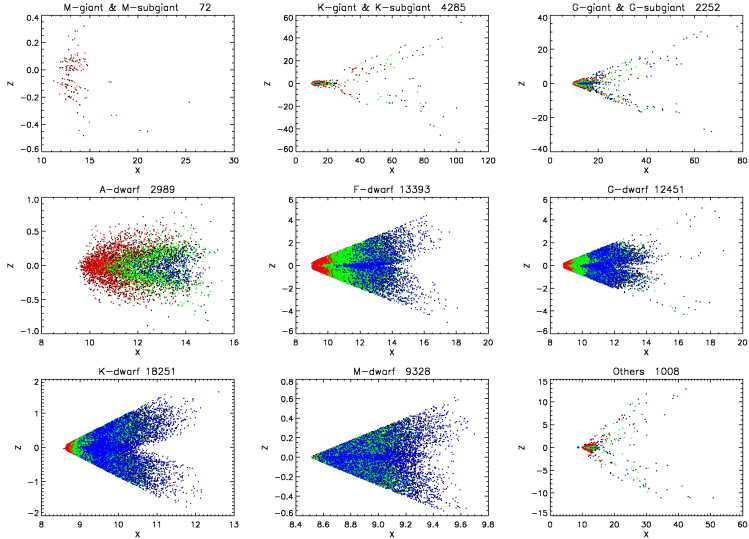} 

 \caption{Spatial distributions of the simulated LSS-GAC targets in the ($X$,
$Z$) plane of Galactic radius and height for a one-degree wide stripe of $|b|
\leq 30^{\circ}$ centered at $l = 180^{\circ}$. The numbers of different types
of star targeted in this stripe are labeled.  Red, green and blue dots
represent sources of B, M and F plates, respectively (c.f. Fig.\,2).}

   \label{NNN:fig3} 
\end{center}
\end{figure}

Once the values of $m_1$ and $m_2$ are set, stars of B, M and F plates are
selected, in batches of 200 stars per sq.deg. For each category, up to
10 batches can be selected pending on the availability of sources. The first
batch of stars are assigned the highest observational priority, the second one
priority lower, and so forth. In selecting each batch of stars, if a star of a
given coordinate set, (RA, Dec) or ($l$, $b$), is selected, then stars in its
vicinity of 2\,arcmin are removed from the pool. Those stars are however put back
to the pool when selecting the next batch of stars.

For each selected field center (c.f. \S{2}) and plate category (B, M or F),
fibers are assigned to stars selected above according to their priorities by
running through SSS, the LAMOST Survey Strategy System, assuming a nominal
observing date and time. Fibers targeting blank sky for the measurements of sky
background, typically 320 per plate, are also allocated.  Excluding dead fibers
($\sim 100$), generally $\sim 3,500$ fibers are available for the LSS-GAC
targets per plate. The SSS has to be rerun near the actual observing date, to
allow for the time-dependence of transformation between the equatorial and
focal plane coordinate systems and to ensure enough guiding stars are
accessible. The changes of stars that eventually get allocated a fiber and
observed are generally small, typically on the level of a few tens or less.

The above survey strategy and target selection algorithm are tested using mock
catalogs generated from the Besan\c{c}on Galactic model (Robin et al. 2003).
As an example, for a $1^{\circ}$ wide stripe of 56.7\,sq.deg.\ centered at $l =
180^{\circ}$ and stretching from $b = -30$ to $+30^{\circ}$, the model yields
396,075 stars of $14 < r \leq 18.5$\,mag. Amongst them, 0.018, 2.3 and 2.1\%
are M, K and G subgiants/giants, respectively, and 1.9, 24, 39, 24 and 4.3\%
are respectively A, F, G, K and M dwarfs. Running this mock catalog through the
above target selection algorithm, one finds that in total 64,029 of all stars,
99, 47 and 27\% of all M, K and G subgiants/giants, 39, 14, 8, 19 and 55\% of
all A, F, G, K and M dwarfs, will be selected and targeted by the LSS-GAC,
respectively. Fig.\,3 shows the spatial distributions of various types of star
targeted in the ($X$, $Z$) plane of Galactic radius and height.  

\section{Survey status}

\begin{figure}[t]
\begin{center}
 \includegraphics[width=4in]{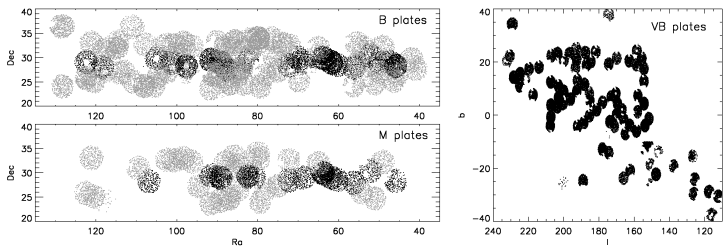} 

  \caption{Spatial distributions of targets from B (top left), M (bottom left)
and VB plates (right). By March 8, 2013, in total 257,317, 115,054, 25,480 and
393,791 spectra of ${\rm S/N}(7450\,{\rm \AA}) \geq 10$ per pixel have been
collected from B, M, F and VB plates, respectively. For B and M plates, black
and grey dots show respectively data collected during the pilot and formal
surveys. Only one-in-ten stars are shown.}

   \label{NNN:fig4} 
\end{center}
\end{figure}

After the completion of construction in June 2009, LAMOST went through a
two-year period of commissioning and performance characterization and
optimization. The problem of fiber-positioning was solved using an innovative
technique. The 4,000 fibers are divided into 160 groups. For each group, the 25
adjacent fibers are utilized to target 25 bright stars, with predefined offsets
following a given pattern. The measured throughput ratios of stellar and sky
light of the 25 fibers are then used to estimate the average fiber positioning
error of this group of fibers in the focal plane. With this method, the
distribution of fiber positioning errors can be mapped out with a single
exposure, even under non-photometric conditions (Yuan et al., in preparation).
Slit masks of 2/3 the fiber diameter were also introduced, increasing the
spectral resolution from $\sim 1,200$ to 1,800, comparable to that of SDSS, at
the expenses of 20\% loss of light.

Following the commissioning period, a pilot survey of one year was initiated in
the fall of 2011, to further test and optimize the survey strategy.  Constant
monitoring of the LAMOST performance shows that for large scale surveys, a
limiting magnitude of 17.8 is probably realistic. Under exceptional conditions
(a few nights per months), a depth of 18.5\,mag. can be reached for limited
number of fields.

The LAMOST formal surveys were initiated in the fall of 2012. Hitherto, within
the LSS-GAC footprint, $\sim 0.15$ and 0.25\,M spectra of ${\rm
S/N}(\lambda7450) \ge 10$ per pixel ($\sim 1.7$\,{\AA}) have been collected
during the periods of pilot and formal surveys, respectively. In addition,
$\sim 0.4$\,M spectra from the VB plates (some outside the LSS-GAC footprint)
have been obtained. The spatial distributions of those sources are shown in
Fig.\,4. 

\section{Data reduction}

The LAMOST 2D, 1D and stellar parameter pipelines have been developed at the
LAMOST Operation and Development Center (Luo et al., this volume). There are
however issues specific to the LSS-GAC, and a set of pipelines, LSP3, tailored
to the need of LSS-GAC have been developed at Peking University (PKU).

Owing to high dust reddening in the disk ($E_{B - V} \sim 0.3$ towards Galactic
anti-center sightlines) flux calibration standards, such as F turnoff stars
adopted by the SDSS, cannot be reliably identified and selected from the
photometric colors alone. On the other hand, stars of all types and colors are
targeted by the LSS-GAC. LSS-GAC spectra are first processed using nominal
spectral response curves (SRCs). The initial stellar parameters are then
derived using the LSP3. Based on the results, 5 -- 10 F stars [$5,500 < T_{\rm
eff} < 7,000$~K and $\log\,g\,{\rm (cm\,s^{-2}}) > 3.0$] per spectrograph are
selected as the flux calibration standards. Extinctions towards those stars are
deduced by comparing the photometric colors to those calculated from the Kurucz
low-resolution model spectra (Castelli \& Kurucz 2004) of the same stellar
parameters. The SRCs are then derived by comparing the observed spectra with
the synthetic spectra of Munari et al. (2005), interpolated to the
corresponding stellar parameters and degraded to the LAMOST spectral
resolution, and then reddened assuming the reddening derived above. The
Fitzpatrick (1999) extinction law of $R = 3.1$ is adopted. The spectra are then
reprocessed using the newly deduced SRCs and values of stellar parameters are
updated. The above process is iterated (twice is sufficient) to yield the final
flux-calibrated spectra.

\begin{figure}[t]
\begin{center}
 \includegraphics[width=5.00in]{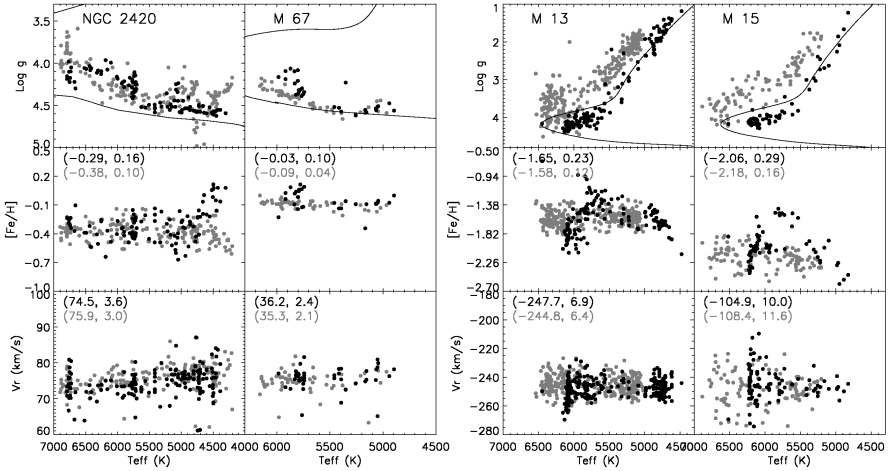} 

  \caption{Comparisons of LSP3 (black dots) and SDSS DR7 (grey dots) stellar
parameters for the open clusters NGC\,2420 and M\,67, and globular clusters
M\,13 and M\,15. For M\,67 and M\,15, a constant value of +40 and
$-140$\,km\,s$^{-1}$ has been added to the measured values, respectively.  Also
overplotted in the top panels are isochrones of (${\rm [Fe/H]}$, ${\rm
[\alpha/Fe]}$, ${\rm age}$) = ($-0.3$, 0, 1\,Gyr) for NGC\,2420, (0, 0,
2.5\,Gyr) for M\,67, ($-1.5$, 0.4, 11.5\,Gyr) for M\,13 and ($-2.0$, 0.4,
12\,Gyr) for M\,15. The clustering of LSP3 points at high temperatures are due
to the scarceness of MILES stars at those temperatures, particularly at low
metallicities.}

   \label{NNN:fig5} 
\end{center}
\end{figure}

As a test of the accuracy of our flux-calibration procedure, colors are
calculated from the calibrated spectra and compared to the photometric values.
For spectra of ${\rm S/N}(\lambda4650) > 30$ per pixel ($\sim 1$~{\AA}), we
find $\Delta(g - r) = 0.02\pm 0.06$\,mag on average. For $(r - i)$, the spectra
yield colors about $0.06\pm 0.04$\,mag. bluer than the photometric values,
owing to the fact that the LSP3 has opted to not to correct for the telluric
absorption bands, most notably in the $i$-band. Comparisons of the LAMOST and
SDSS DR7 spectra of common objects show that for high extinction regions, the
LSP3 yields more realistic spectral energy distributions (SEDs) than the SDSS.
Finally, comparisons of the spectra obtained at different epochs indicate a
calibration accuracy of better than 8\% for the wavelength range
$\lambda\lambda$4000--9000. The SRCs are found to vary with time, by as much as
20\% in a given night and even larger for different nights. The reasons are
unclear. For the moment, it seems essential to derive SRCs for each observed
plate.

The LSP3 derives stellar parameters by matching the observed spectra to the
empirical spectral library MILES (S\'{a}nchez-Bl\'{a}zquez et al. 2006).
Compared to the ELODIE library (Prugniel \& Soubiran 2001), the MILES library
has the advantage that they are obtained with a spectral resolution
(2.3\,{\AA}), comparable to that of LAMOST, and carefully flux-calibrated. Both
libraries have similar numbers of stars and parameter space coverage, with the
parameters determined from high resolution spectroscopy. For radial velocity
determinations, the ELODIE 3.1 library is used. 
 
To obtain the initial values of radial velocity $V_{\rm r}$ and stellar
parameters $T_{\rm eff}$, $\log\,g$ and [Fe/H] of a target spectrum, we first
normalize the spectrum in a way similar to the SSPP (Lee et al. 2008), using a
ninth and fourth order polynomial for the blue- ($\lambda\lambda$3800--5800)
and red-arm ($\lambda\lambda$6100--9000) spectra, respectively, and compare the
results to the normalized MILES spectra. The biweight means and standard
deviations of stellar parameters of the best 20 matches are then adopted as the
initial parameters of the target spectrum. The initial $V_{\rm r}$ is then
obtained by cross-correlating with the best match.

For further iterations, we first fix $V_{\rm r}$ and select a 3D box in the
$T_{\rm eff}$, $\log\,g$ and [Fe/H] parameter space centered at the initial
values. The sizes of box are set to 3 times the standard deviations of initial
values or values $0.2\times T_{\rm eff}$ in $T_{\rm eff}$, 1.0\,dex in
$\log\,g$ and 1.0\,dex in [Fe/H], whichever is bigger for each of three
quantities. The observed target spectrum, un-normalized, is then matched with
the MILES spectra, again un-normalized, in the box. To account for the possible
effects of interstellar extinction and uncertainties in flux-calibration, a
third order polynomial is allowed to scale the SEDs of MILES spectra to that of
the target. The optimization is carried out separately for two wavelength
ranges, $\lambda\lambda$4000--5500 and $\lambda\lambda$6100--6800, using both
the minimum $\chi^2$ and maximum cross-correlation techniques. Strong city
light emission lines and night sky emission lines, e.g. the Hg~{\sc i}
$\lambda$4358 and [O~{\sc i}] $\lambda\lambda$6300,6363, are masked out. With
the newly determined stellar parameters, $V_{\rm r}$ is then redetermined using
the ELODIE library. The process is repeated until $V_{\rm r}$ varies less than
3.0\,km\,s$^{-1}$ in two consecutive iterations. In most cases, the results
converge in one or two iterations. The biweight means of stellar parameters of
the four best templates in the final iteration are then adopted as the final
values of the target spectrum. Currently, only results from the blue wavelength
range are used.
 
\begin{figure}[t] \begin{center}
\includegraphics[width=4in]{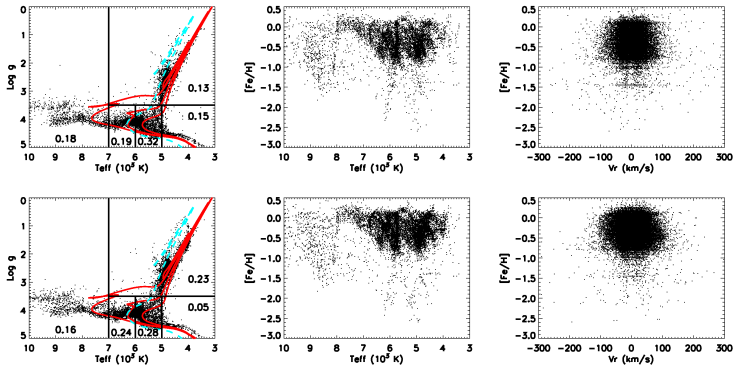} 

  \caption{Radial velocities and stellar parameters determined from 197,821
spectra of B, M and F plates (top panels) and 393,791 spectra of VB plates
(bottom panels) of ${\rm S/N}(\lambda4650) > 10$ per pixel. Only one-in-ten
stars are shown. In the left two panels, the cyan dashed line shows the
isochrone of metallicity ${\rm [Fe/H]} = -1$ and age 10\,Gyr, while the three
red lines, from left to right, show the isochrones of ${\rm [Fe/H]} = 0$ and
ages 1, 3 and 10\,Gyr, respectively. The $T_{\rm eff}$ -- $\log\,g$ plane is
divided into different regions. The fractions of stars of individual regions
are labeled.}

   \label{NNN:fig6} 
\end{center}
\end{figure}

Comparisons of multi-epoch observations, and with the results of RAVE
(Steinmetz et al. 2006) and SDSS show that for F, G and K stars, the LSP3 has
achieved an accuracy of 8 and 5\,km\,s$^{-1}$ for ${\rm S/N}(\lambda4650) > 10$
and 20, respectively. For hotter stars, the uncertainties are $\sim 20$ and
7\,km\,s$^{-1}$ for ${\rm S/N}(\lambda4650) \gtrsim 10$ and 20, respectively.
For stellar parameters, the uncertainties are about 110\,K, 0.15\,dex and
0.15\,dex for $T_{\rm eff}$, $\log\,g$ and [Fe/H], respectively.  The stellar
parameters yielded by LSP3 agree well with those given by ULySS (Koleva et al.
2009; Wu et al. 2012), a pipeline based on the ELODIE library. Fig.\,5 compares
the LSP3 radial velocities and stellar parameters determined from SDSS spectra
with those of SDSS DR7 for member stars of two open clusters and two globular
clusters.  Some artifacts are clearly visible in the distributions of LSP3
parameters, owing to incomplete or sparse parameter coverage of the MILES
stars, in particular at low metallicities. Efforts are underway to fill the
gaps in parameter space of the MILES library by obtaining additional spectra
using the NAOC 2.16\,m telescope. 

Fig.\,6 shows distributions of radial velocities and stellar parameters
determined for 197,821 spectra of ${\rm S/N}(\lambda4650) > 10$ per pixel
obtained with B, M and F plates, and for 393,791 spectra obtained with VB
plates. Artifacts apparent in the SDSS DR7 and DR9 releases, e.g.  two false
branches of stars in the $T_{\rm eff}$ -- $\log\,g$ HR diagram, one near the
turn-off stars ($T_{\rm eff} \sim 6,200$~K) and another at low temperatures
($4,000 \lesssim T_{\rm eff} \lesssim 4,800$~K), are no longer present. The
LSP3 also seems to yield stellar parameters in better agreement with the
theoretical isochrones. 

\section{Examples of application}

\begin{figure}[t]
\begin{center}
 \includegraphics[width=5.25in]{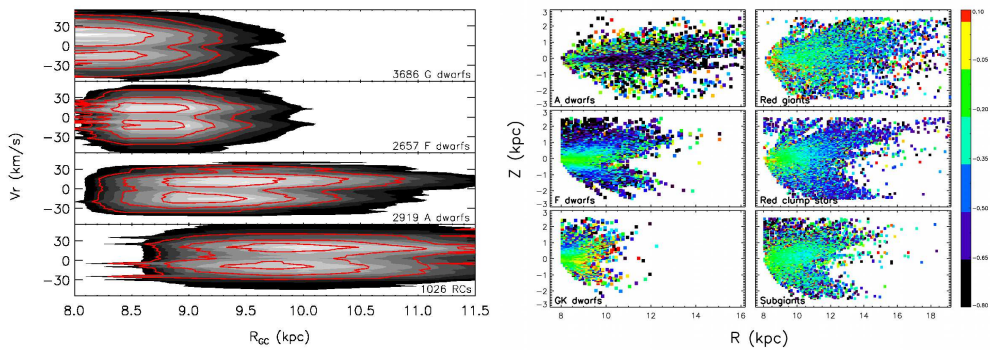} 

  \caption{{\it Left:}\ Bifurcation of radial velocities of disk G, F and A
dwarfs and RCs of heights $|Z| < 0.1$\,kpc in the direction of Galactic
anti-center ($178 < l < 182^{\circ}$).  {\it Right:}\ Metallicity distributions
traced by different types of star.}

   \label{NNN:fig7} 
\end{center}
\end{figure}

Although the LSS-GAC survey and data reduction are still in early stage, nearly
one million spectra and resultant radial velocities, stellar parameters are
already available for scientific exploitation. Taking advantage of the large
number of spectra offered by the SDSS/SEGUE and using an innovative technique
of paring stars of high extinctions with those of very low or nil extinctions
but otherwise of identical stellar parameters, Yuan \& Liu (2012) report the
detections of diffuse interstellar bands towards thousands of sightlines of
Galactic stars. The same technique has been utilized to derive the extinction
coefficients and study the extinction law towards thousands of sightlines
(Yuan, Liu \& Xiang 2013).  Applying this technique to the LSS-GAC data, Yuan
et al. (this volume) has produced 3D extinction maps of the outer parts of
Galactic disk. In spite of the incomplete sky coverage, distinct features, such
as the Perseus and Outer Arms, and effects of warps of the outer disk, are
clearly visible in the deduced extinction maps. In the work, once the
extinction towards the sightline of a given star has been determined by
combining the XSTPS-GAC and 2MASS photometry, given the $T_{\rm eff}$,
$\log\,g$ and [Fe/H] parameters yielded by LSP3, the star's distance is then
estimated by finding the closest model in the Dartmouth Stellar Evolution
Database (Dotter et al. 2008) except for red clump stars (RCs).  Reliability
and accuracy of the algorithm are tested using multiple observations, star
clusters and the Hipparcos distances of stars in the MILES and ELODIE
libraries. We find that the algorithm yields distances accurate to $\sim 15$
and 30\% for dwarfs and giants, respectively, when the spectra have ${\rm
S/N}(\lambda4650) \ge 15$. However, we also find that the distances thus
derived are systematically smaller than the Hipparcos values by respectively
$\sim 6$ and 12\% for dwarfs and giants. RCs are good standard candles and
easily identifiable in the $T_{\rm eff}$ -- $\log\,g$ HR diagram, falling in
the region of $4,500 < T_{\rm eff} < 5,200$\,K and $2 < \log\,g\,{\rm
(cm\,s^{-2})} < 3$. Their distances are calculated assuming absolute
magnitudes $M_i = 0.189$ and $M_{Ks} = -1.567$ (Gronewegen 2008), with an
estimated accuracy of $\sim 10$\%.  Currently, a Bayesian approach as proposed
by Burnett \& Binney (2010) is being considered for implementation in order to
determine stellar distances, ages, masses and metallicities simultaneously.

Liu et al. (2012) find that the radial velocities of a sample of 697 RCs
observed with the MMT 6.5\,m telescope show a bifurcation distribution split by
$~30$\,km\,s$^{-1}$, peaking at Galactic radius of $\sim 10$\,kpc, in
coincidence with the location of the Perseus arm, and interpret the result as
evidence of the presence of the outer Lindblad resonance of the Galactic bar,
although the possibility of the corotation resonance of the spiral arms can not
be ruled out. The left panels of Fig.\,7 show the radial velocity distributions
of thousands of disk G, F, A dwarfs and RCs from the LSS-GAC survey in the
direction of Galactic anti-center. Bifurcation is clearly seen in all cases,
peaking at different radii, from the solar neighborhood to $\sim 11$\,kpc, in
line with the expected survey depths of individual types of tracer of concern
here.  In fact, a bifurcation distribution is seen in all directions, from
longitudes 150 to $210^{\circ}$ and of Galactic heights $|Z| \leq 0.8$\,kpc,
casting doubt upon a resonant interpretation. 

The right panels of Fig.\,7 show metallicity distributions traced by different
types of star. While the data including distances need to be carefully checked,
they have already shown some interesting trends, e.g.  significant vertical but
marginal radial gradients. It is also interesting to note that RCs seem to
indicate a clear radial gradient beyond 10\,kpc.

In conclusion, the LSS-GAC is well under way and poses to make major
contributions to our understanding of the stellar populations, kinematics, and
the chemical enrichment and star formation history of the Milky Way.

\begin{discussion}

\discuss{Nissen}{You mentioned the possibility to measure [$\alpha$/Fe]
and [C/Fe]. Do you have any results on that already?}

\discuss{Liu}{Not yet. The possibility is demonstrated by Marstellar et al.
(2009, AJ, 138, 533) and Lee et al. (2011, AJ, 141, 90), although we believe
the methods need careful calibration.}

\discuss{Binney}{Are you delivering distances in parallel with the other
parameters? Can you say something about the accuracy of your distances?}

\discuss{Liu}{Yes, extinctions and distances will be released along with the
LSP3 parameters, and can be obtained from the author (XWL) upon request,
subject to regulations of the LAMOST data policy. A brief discussion of the
accuracy is presented in \S{6}.} 

\discuss{Ludwig}{What is the data policy of the LSS-GAC survey? Will and if so
when will the data be released to the community?}

\discuss{Liu}{LSS-GAC is parts of the LAMOST surveys and follows the same data
policy.}

\end{discussion}

\end{document}